\documentclass{aa}
\let\tmp\newinsert
\let\newinsert\newbox
\usepackage{floatrow}
\let\newinsert\tmp
\usepackage[varg]{txfonts}
\usepackage{natbib}
\bibpunct{(}{)}{;}{a}{}{,} % to follow the A&A style
\usepackage{hyperref}
\usepackage{enumitem}% http://ctan.org/pkg/enumitem
\usepackage{amsmath}
\usepackage[dvipsnames]{xcolor}
\usepackage{subcaption}
\usepackage{graphicx,epstopdf}
\usepackage{siunitx} %for si units
\newcommand{\mg}[1]{#1}
\newcommand{\gm}[1]{#1}
\usepackage{makecell} % for multiple lines in table with \thead
\usepackage{amssymb} %for mathematical symbols
\usepackage[font=small,labelfont=bf,labelsep=space]{caption}
\usepackage{enumitem}% http://ctan.org/pkg/enumitem%
\usepackage{etoolbox}
\usepackage{tikz}
\newrobustcmd*{\mysquare}[1]{\tikz{\filldraw[draw=#1,fill=#1] (0,0) rectangle (0.18cm,0.18cm);}}
\newrobustcmd*{\mycircle}[1]{\tikz{\draw[draw=#1] (0,0) circle [radius=0.1cm];}}
\newrobustcmd*{\mybullet}[1]{\tikz{\filldraw[draw=#1,fill=#1] (0,0) circle [radius=0.1cm];}}
\DeclareMathAlphabet{\mathitbf}{OML}{cmm}{b}{it}
\newcommand{\Avec}{\mathitbf{A}}

\newcommand{\Apvec}{\mathitbf{A}_{\mathrm{0}}}
\newcommand{\Bvec}{\mathitbf{B}}

\newcommand{\Bpvec}{\mathitbf{B}_{\mathrm{0}}}

\newcommand{\Jvec}{\mathitbf{J}}

\newcommand{\flux}{\phi}
\newcommand{\Etot}{E}
\newcommand{\Epot}{E_{\mathrm{0}}}
\newcommand{\Efree}{E_{\mathrm{F}}}
\newcommand{\Efprime}{\Efree/\Etot}
\newcommand{\Ediv}{E_{\mathrm{div}}}
\newcommand{\Ejs}{E_{\mathrm{J,s}}}

\newcommand{\Eps}{E_{\mathrm{0,s}}}
\newcommand{\Ej}{E_{\mathrm{J}}}
\newcommand{\Ejns}{E_{\mathrm{J,ns}}}
\newcommand{\Epns}{E_{\mathrm{0,ns}}}
\newcommand{\Emix}{E_{\mathrm{mix}}}
\newcommand{\Edivprime}{E_{\rm div}/E}

\newcommand{\thetaj}{\theta_J}

\newcommand{\hpj}{H_{\mathrm{PJ}}}
\newcommand{\hj}{H_{\mathrm{J}}}
\newcommand{\hjprime}{|\hj|/|\hv|}
\newcommand{\hv}{H_{V}}

\newcommand{\hvnorm}{|\hv|/{\phi^{\prime}}^{2}}
\newcommand{\hjnorm}{|\hj|/{\phi^{\prime}}^{2}}

\newcommand{\eg}{{\it e.g.}}
\newcommand{\fluxmean}{\langle{\flux}\rangle}
\newcommand{\hvmean}{\langle{H}_{V}\rangle}
\newcommand{\hjmean}{\langle{H}_{\mathrm{J}}\rangle}
\newcommand{\hpjmean}{\langle{H}_{\mathrm{PJ}}\rangle}
\newcommand{\Etotmean}{\langle{E}\rangle}
\newcommand{\Efreemean}{\langle{E}_{\mathrm{F}}\rangle}
\newcommand{\hjprimemean}{\langle{\hjprime}\rangle}
\newcommand{\Efprimemean}{\langle{\Efprime}\rangle}
\newcommand{\hvnormmean}{\langle{\hvnorm}\rangle}
\newcommand{\hjnormmean}{\langle{\hjnorm}\rangle}
\newcommand{\myplots}[4]{
    \hspace{2em}\begin{subfigure}[t]{0.33\textwidth}
        \centering
        \includegraphics[width=1.0\linewidth]{plots/#3_mu3_1e-3#1Ediv_E_lim_0.1CW_thetalim_30#4}\\
    \end{subfigure}\hspace{3em}
     \vspace{-0.75em}
    \begin{subfigure}[t]{0.33\textwidth}
        \centering
        \includegraphics[width=1.0\linewidth]{plots/#3_mu3_1e-3#2Ediv_E_lim_0.1CW_thetalim_30#4}\\
    \end{subfigure}\hspace{5em}  
    
}
\newcommand{\myplotsapp}[4]{
    \hspace{2em}\begin{subfigure}[t]{0.35\textwidth}
        \centering
        \includegraphics[width=1.0\linewidth]{plots/#3_mu3_1e-3#1Ediv_E_lim_0.1CW_thetalim_30#4}\\
    \end{subfigure}\hspace{3em}
     \vspace{-0.7em}
    \begin{subfigure}[t]{0.35\textwidth}
        \centering
        \includegraphics[width=1.0\linewidth]{plots/#3_mu3_1e-3#2Ediv_E_lim_0.1CW_thetalim_30#4}\\
    \end{subfigure}\hspace{5em}  
    
}
\newcommand{\jktplot}[4]{
    \hspace{2em}\begin{subfigure}[t]{0.36\textwidth}
        \centering
        \includegraphics[width=1.0\linewidth]{plots/#3_mu3_1e-3#1Ediv_E_lim_0.1CW_thetalim_30#4}\\
    \end{subfigure}\hspace{2em}
     \vspace{-0.59em}
    \begin{subfigure}[t]{0.36\textwidth}
        \centering
        \includegraphics[width=1.0\linewidth]{plots/#3_mu3_1e-3#2Ediv_E_lim_0.1CW_thetalim_30#4}\\
    \end{subfigure}\hspace{5em}  
    
}

\newcommand{\fig}[2]{
    \myplots{12673}{12192}{#1}{#2}
    \myplots{11158}{12268}{#1}{#2}
    \myplots{11429}{11302}{#1}{#2}
    \myplots{12297}{11339}{#1}{#2}
    \myplots{11890}{11166}{#1}{#2}
}
\newcommand{\appfig}[2]{
    \myplotsapp{12673}{12192}{#1}{#2}
    \myplotsapp{11158}{12268}{#1}{#2}
    \myplotsapp{11429}{11302}{#1}{#2}
    \myplotsapp{12297}{11339}{#1}{#2}
    \myplotsapp{11890}{11166}{#1}{#2}
}
\newcommand{\jktfig}[2]{
    \jktplot{12673}{12192}{#1}{#2}
    \jktplot{11158}{12268}{#1}{#2}
    \jktplot{11429}{11302}{#1}{#2}
    \jktplot{12297}{11339}{#1}{#2}
    \jktplot{11890}{11166}{#1}{#2}
}
\title{Magnetic helicity and energy budget around large confined and eruptive solar flares}
\author{Manu Gupta\inst{\ref{inst1}} \and J. K. Thalmann\inst{\ref{inst1}}\and A. M. Veronig\inst{\ref{inst1},\ref{inst2}}}

\institute{%
University of Graz, Institute of Physics/IGAM, Universit\"atsplatz 5, 8010 Graz, Austria\label{inst1} 
\email{manu.gupta@uni-graz.at} \and Kanzelhöhe Observatory for Solar and Environmental Research, University of Graz, Austria\label{inst2}
}
\date{Received 17 February 2021 / Accepted 9 June 2021 }
\abstract{
{\textit{Context:} 
In order to better understand the underlying process and prerequisites for solar activity, it is essential to study the time evolution of the coronal magnetic field of solar active regions (ARs) associated to flare activity.
} \\
{\textit{Aims:} We investigate the coronal magnetic energy and helicity budgets of ten solar ARs, around the times of large flares. In particular, we are interested in a possible relation of the derived quantities to the particular type of the flares that the AR produces, i.e., whether they are associated with a CME or \gm{they are confined}.
} \\
{\textit{Methods:} Using an optimization approach, we employ time series of 3D nonlinear force-free magnetic field models of ten ARs, covering a time span of several hours around the time of occurrence of large solar flares (GOES class M1.0 and larger). We subsequently compute the 3D magnetic vector potentials associated to the model 3D coronal magnetic field using a finite-volume method. This allows us to correspondingly compute the coronal magnetic energy and helicity budgets, as well as related (intensive) quantities such as the relative contribution of free magnetic energy, $\Efprime$ (energy ratio), the fraction of non-potential (current-carrying) helicity, $\hjprime$ (helicity ratio), and the normalized current-carrying helicity, $\hjnorm$.
} \\
{\textit{Results:} The total energy and helicity budgets of flare-productive ARs (extensive parameters) cover a broad range of magnitudes, with no obvious relation to the eruptive potential of the individual ARs, i.e., whether or not a CME is produced in association with the flare. The intensive eruptivity proxies, $\Efprime$ and $\hjprime$, and $\hjnorm$, however, seem to be distinctly different for ARs that produced CME-associated large flares compared to those which produced confined flares. For the majority of ARs in our sample, we are able to identify characteristic pre-flare magnitudes of the intensive quantities, clearly associated to subsequent CME-productivity.\\
}
{\textit{Conclusions:} If the corona of an AR exhibits characteristic values of $\hjprimemean>0.1$, $\Efprimemean>0.2$, and $\hjnormmean>0.005$, then the AR is likely to produce large CME-associated flares. Conversely, confined large flares tend to originate from ARs that exhibit coronal values of $\hjprimemean\lesssim0.1$, $\Efprimemean\lesssim0.1$, and $\hjnormmean\lesssim0.002$.
} 
}


\begin{document}

\titlerunning{Magnetic helicity and energy budget around large confined and eruptive solar flares}
%\authorrunning{}
\maketitle

\section{Introduction}\label{sec:introduction}
Solar flares are powerful explosive events observed in the solar atmosphere that often occur in association with coronal mass ejections (CMEs). During flares, a vast amount of magnetic energy is released, converted into particle acceleration, kinetic energy of mass ejections, plasma heating, and leading to enhanced emission of radiation across the electromagnetic spectrum \citep[see reviews by][]{2011SSRv..159...19F,2017LRSP...14....2B}. The underlying physical process involves accumulation of free magnetic energy in the corona by induced electric currents, which is eventually released by magnetic reconnection \citep[for a review see, e.g.,][]{2002A&ARv..10..313P,2011LRSP....8....6S}. While smaller flare events tend to be confined in nature (\gm{i.e.,} are not accompanied by a CME), large flares mostly show an eruptive character \citep[i.e., they are associated with a CME; see, e.g.,][]{1992LNP...399....1S,2003SoPh..218..261A,2005JGRA..11012S05Y}.
CMEs carry plasma and embedded magnetic field into the interplanetary space and can cause major geomagnetic storms upon interaction with Earth\textquotesingle s magnetosphere \citep[e.g.,][]{1988JGR....93.8519T, 2017SSRv..212.1137K}.

Over the last two decades, many attempts have been undertaken to forecast flares. Some of these flare prediction methods rely on the characterization of distinct features at photospheric levels, based on observed magnetic field or intensity images, within the host active region (AR). Other so-called statistics-based approaches utilize information on prior flare activity only. \cite{2016ApJ...829...89B} systematically compared such flare prediction methods and showed that the different approaches have a comparable flare prediction ability but require further improvement. \cite{2003ApJ...595.1296L,2007ApJ...656.1173L} concluded in their study that the information contained in the photospheric magnetic field at any particular time is limited. This is because the parameters deduced from the 2D imprint of the intrinsically 3D coronal magnetic field, underestimate essential aspects which contribute to a
flare process \citep[see also, e.g.,][]{Barnes_2006}.

Significant progress has also been made in understanding the physical requirements for the occurrence of CMEs. Different approaches have been developed based on observations or modeling to assess whether a flare is likely to be eruptive (i.e., CME-associated) or confined \citep[for a review on CME prediction methods, see \gm{Sect.} 3 in][]{2018SSRv..214...46G}. \cite{Wang_2007} examined 14 X-class flares and found that the location of a flare within the host AR plays a significant role. In particular, confined large flares tend to originate from locations closer to the AR center, while eruptive flares more likely originate from the AR periphery. \cite{2018EGUGA..20.5038B} analyzed 44 flares of GOES class $\geqslant$ M5.0 and also noted that confined large flares primarily originate from locations close to the flux-weighted center of the ARs. In contrast, flares tend to be eruptive if originating from compact ARs (i.e., ARs whose characteristic distance between the flux-weighted magnetic polarity centers is $\lesssim60$~Mm) or if they originate from the periphery of ARs.

Concerning the coronal pre-requisites for flare productivity, observation-based 3D modeling of the coronal magnetic field delivered important insights. The free magnetic energy previously stored in the AR corona appears to be related to the frequency and size (magnitude) of upcoming flares \citep[e.g.,][]{2010ApJ...713..440J,2014ApJ...788..150S}. A certain pre-flare free magnetic energy content,
however, is not decisive whether a flare will occur, as not necessarily all of the coronal free
magnetic energy is released at once during a flare \citep[e.g.,][]{2012SoPh..276..133G}. These findings suggest that dissipative quantities, such as the magnetic flux or the free magnetic energy, have a limited ability for flare prediction, as they are not uniquely related to the complexity of the flare-involved magnetic field. In this respect, magnetic helicity has a number of advantageous properties, as it depends not only on the strength but is also unique to the structure of the underlying magnetic field. In addition, contrary to magnetic energy, it is very well conserved even in non-ideal dynamics \cite{1984GApFD..30...79B}.

\cite{2012ApJ...759L...4T} found a robust correlation between the free magnetic energy and the magnetic helicity in flaring ARs, and suggested that magnetic helicity represent an essential ingredient in addition to free energy for large solar eruptions. \cite{2008ApJ...686.1397P} noted that the coronal magnetic helicity accumulates at a nearly constant rate, and stagnates prior to the onset of individual flares. Therefore, studying the time variations of the magnetic helicity and the free energy can help us to better understand the evolution of the active region corona, and the process leading to a flare and CME. 

Magnetic helicity is a measure of geometrical complexity of the magnetic fields, summing the individual contributions of twist and writhe, as well as the interlinking of magnetic structures \citep{1969JFM....35..117M}. It is a conserved quantity of ideal magneto-hydrodynamics (MHD), and almost conserved in resistive MHD (\cite{1974PhRvL..33.1139T}; \cite{1984JFM...147..133B}; \cite{2015A&A...580A.128P}). 
The basic definition of magnetic helicity is gauge invariant only for magnetically closed systems, i.e., where no field threads the considered volume's boundary. For applications to open systems, such as the solar corona, \cite{1984JFM...147..133B} and \cite{finn1985magnetic} formulated a relative measure for the total helicity within a considered volume, $V$, in the form,
\begin{equation}
	\hv=\int_V\left(\Avec+\Apvec\right)\cdot\left(\Bvec-\Bpvec\right) {\rm ~d}V, \label{eq:hv}
\end{equation}
where $\Bvec$ is the field under study, and $\Bpvec$ is the reference field that is usually assumed to be potential (current-free) and shares the normal component with $\Bvec$ on the boundary of $V$. Furthermore, $\Avec$ and $\Apvec$ are the respective vector potentials satisfying $\Bvec=\nabla\times\Avec$ and $\Bpvec=\nabla\times\Apvec$. The  gauge-invariant formulation of $\hv$ in \href{eq:hv}{\gm{Eq.}~(\ref{eq:hv})} is conserved, both in ideal and resistive MHD \citep{2018ApJ...865...52L}, but is not additive \citep{2020A&A...643A..26V}.

\href{eq:hv}{Equation~(\ref{eq:hv})} can be decomposed as, $\hv=\hj+\hpj$ \citep{1999PPCF...41B.167B, 2003and..book..345B}, with
\begin{eqnarray}
	\hj&=&\int_V\left(\Avec-\Apvec\right)\cdot\left(\Bvec-\Bpvec\right) {\rm ~d}V, \label{eq:hj}
\end{eqnarray}
and,
\begin{eqnarray}
	\hpj&=&2\int_V\Apvec\cdot\left(\Bvec-\Bpvec\right) {\rm ~d}V, \label{eq:hpj}
\end{eqnarray}
where $\hj$ is the magnetic helicity of the current-carrying (non-potential) component of the magnetic field, and $\hpj$ is the volume-threading helicity between $\Bpvec$ and $\Bvec$. Both $\hj$ and $\hpj$ are separately gauge invariant but are not individually conserved quantities neither in ideal nor resistive MHD \citep{2018ApJ...865...52L}.

In recent years, a number of studies have been carried out to comprehensively study the relative magnetic helicity and its individual contributors ($\hj$ and $\hpj$) using MHD simulations. \cite{2017A&A...601A.125P} studied seven different 3D visco-resistive MHD simulations of the emergence of a twisted magnetic flux rope into a stratified atmosphere. In their study, they observed higher values of $\hj$ during the pre-eruptive phase in those simulation runs which produced a CME-like ejection. Another major finding of their study was that the ratio of the current-carrying helicity to the total helicity, $\hjprime$ ("helicity ratio"), discriminates eruptive from non-eruptive cases already at very early stages of the simulations. Moreover, the values of the helicity ratio were several times higher for the eruptive simulations in comparison to the non-eruptive simulations, with $\hjprime$ $\gtrsim0.45$ before the model eruption. \cite{2018ApJ...863...41Z} investigated five different line-tied MHD simulations involving torus instability as a trigger for eruption. They noted a common value of $\hjprime\simeq0.3$ prior to the onset of the torus instability, for all eruptive simulations. Such high values were not observed for any other investigated quantity. 
 
Also, recent observation-based model works suggest that the helicity ratio has a great ability to characterize the eruptive potential also of solar ARs. \cite{2018ApJ...855L..16J} were the first to provide an observation-based estimate of the helicity ratio, based on a 3D nonlinear force-free (NLFF) model of the AR 11504. They found a value of $\hjprime\simeq0.17$, one hour prior to the onset of an eruptive flare. \cite{2019ApJ...887...64T} studied the time evolution of $\hjprime$ during the disk passage of AR 11158 and AR 12192, based on time series of NLFF models. They reported values of $\hjprime\gtrsim0.17$ before major eruptive flares that originated from AR 11158, and lower values prior to the large confined flares that originated from AR 12192. In recent follow-up studies by \cite{2019A&A...628A..50M} and \cite{2020A&A...643A.153T}, similar values of $\hjprime\gtrsim0.15$ were observed before the eruption of an eruptive X9.3 flare from AR 12673.

Based on these exemplary case studies, it appears that the helicity ratio is a promising candidate to discriminate ARs which produce large eruptive flares from ARs that produce mostly confined flares. This motivates us to perform a first comprehensive study, including a larger number of ARs, and to analyze in detail the time evolution of the coronal magnetic energy and helicity before and around the time of large flares (GOES class M1.0 and larger). In particular, we are interested in the coronal characteristics of the ARs under study, depending on the flare type (eruptive or confined) and investigate the evolution of helicity and energy-based proxies and test their ability to indicate the eruptive potential of solar ARs.

\section{Data and methods}
\subsection{Active region sample}\label{ss:AR}
We select ten solar ARs that produced large flares (GOES class M1.0 and larger). Five of those flare-productive ARs were the source of predominantly CME-associated flares (AR 12673, 11158, 11429, 12297, 11890), whereas the majority of large flares that originated from the other five ARs were confined (AR 12192, 12268, 11302, 11339, 11166). We note here that nine of these selected ARs were analyzed in \cite{2020ApJ...893..123A} regarding their degree of current neutralization, polarity inversion line (PIL), magnetic shear, and the unsigned magnetic flux. 

For each AR, we consider a time interval of 6--10 hours around the largest flare produced by the AR that occurred close to disk center. The corresponding flare type (confined or eruptive) was verified using the LASCO CME catalog \citep{2009EM&P..104..295G} and from Table 1 in \cite{2017SPD....4820001T}. Our AR sample is briefly summarized in the following (and see also \href{t1}{Table~\ref{t1}} for corresponding information).

\begin{itemize}
\item AR 12673  produced two X-class flares on 2017 September 6. The first  one  was a confined X2.2. flare (SOL2017-09-06T08:57). The second flare (SOL2017-09-06T11:53) started 2.6 hours later, and was the largest flare of cycle 24. This eruptive X9.3 flare was accompanied by a fast ($v \approx 1570$ km/s) halo CME according to the SOHO/LASCO CME catalog\footnote{Linear speed of the CME as provided in the SOHO/LASCO CME catalog, \url{https://cdaw.gsfc.nasa.gov/CME_list}.}. This is the only AR in our sample for which two large flares occur within the selected analysis time window of 10 hours.
\item AR 11158 produced an X2.2 flare (SOL2011-02-15T01:\mg{44}) accompanied by a halo CME ($v \approx 670$ km/s). 
\item AR 11429 produced an X5.4 flare (SOL2012-03-07T00:\mg{02}) accompanied by a fast halo CME ($v \approx 2680$ km/s).
\item AR 12297 produced an X2.1 flare (SOL2015-03-11T16:11) that was associated with a slow CME ($v \approx 240$ km/s).
\item AR 11890 produced an X1.1 flare (SOL2013-11-08T04:\mg{20}) with a slow CME ($v \approx 340$ km/s; see the event 25 of Table 1 in \cite{2017SPD....4820001T}, and \cite{2020ApJ...897L..23K}).
\item AR 12192 produced numerous GOES M- and X-class flares, all except one confined. \cite{2015ApJ...801L..23T} and \cite{2015ApJ...804L..28S} suggested, the strong background field in AR 12192 to be the reason for the confinement of the flares that prevented an otherwise unstable flux rope from erupting. Our study here focuses on the confined X3.1 flare (SOL2014-10-24T21:\mg{07}).
\item AR 12268 produced solely confined flares \citep{2019ApJ...871..105Z}. Here, we analyze the coronal evolution around the confined M2.0 flare (SOL2015-01-30T00:32).
\item AR 11302 produced a confined M4.0 flare (SOL2011-09-26T05:06).
\item AR 11339 produced only C- and M-class flares. Here, we focus on the confined M1.8 flare (SOL2011-11-05T20:31).
\item AR 11166 produced a confined X1.5 flare (SOL2011-03-09T23:13).
\end{itemize}

\begin{table}
%\captionabove
\footnotesize
\caption{\label{t1} List of flares under study. For each flare we list the National Oceanic and Atmospheric Administration (NOAA) number of the host AR, the GOES soft X-ray (SXR) class, % studied, CME presence,
the CME association, the heliographic flare location, as well as the solar flare target identifier following the notation of \cite{2010SoPh..263....1L}.}
\centering
\begin{tabular}{lcccc}
\hline\hline
%rule{0pt} to create space after horizontal line
\rule{0pt}{4ex} NOAA & SXR class & CME & Location & Flare target identifier\\
\hline
 12673 &X9.3 &Yes &S09W34 &SOL2017-09-06T11:53\\ %13 \& 9 min 
 11158 &X2.2 &Yes &S20W10 &SOL2011-02-15T01:\mg{44}\\%12 min
 11429 &X5.4 &Yes &N18E31 &SOL2012-03-07T00:\mg{02}\\%22 min
 12297 &X2.1 &Yes &S17E22 &SOL2015-03-11T16:11\\%11 min
 11890 &X1.1 &Yes &S13E13 &SOL2013-11-08T04:\mg{20}\\
\hline
 12192 &X3.1 &No &S22W21 &SOL2014-10-24T21:\mg{07}\\%33 min
 12268 &M2.0 &No &S13W16 &SOL2015-01-30T00:32\\%12 \& 7 min
 11302 &M4.0 &No &N13E34 &SOL2011-09-26T05:06\\%2 min
 11339 &M1.8 &No &N21E37 &SOL2011-11-05T20:31\\%7 min
 11166 &X1.5 &No &N08W11 &SOL2011-03-09T23:13\\
\hline
\end{tabular}
\end{table}
%flare location from: https://www.lmsal.com/solarsoft/latest_events_archive/events_summary

\subsection{Magnetic field modeling and helicity computation}\label{ss:helicity}

The application of \href{eq:hv}{\gm{Eq.}~(\ref{eq:hv})} to the solar corona is hampered by several difficulties, most severely by our inability to measure the coronal magnetic field reliably on a routine basis \citep[e.g., reviews by][]{2009SSRv..144..413C, 2014A&ARv..22...78W}. Therefore, the coronal magnetic field is typically approximated by a nonlinear force-free (NLFF) field within a finite volume, which requires the measured surface vector magnetic field as an input %the lower boundary condition 
\citep[see reviews by][]{2012LRSP....9....5W,2017SSRv..210..249W}. 

The particular approach used in our study is an optimization method, applied to appropriate boundary conditions (deduced from observed photospheric vector magnetic field measurements) to numerically iterate for an approximate %a 
solution to the force-free equations, $\textbf{j} \times \textbf{B}=\bf{0}$ and $\nabla \cdot \textbf{B}=0$, within a chosen %cubic 
computational domain. We apply the method of \cite{2010A&A...516A.107W} using the 12-min cadence {\sc hmi.sharp\_720s} data, constructed from polarization measurements of the Helioseismic and Magnetic Imager \citep[HMI;][]{2012SoPh..275..207S} onboard the Solar Dynamics Observatory \citep[SDO;][]{2012SoPh..275....3P}. The data set is binned by a factor of two to a spatial resolution of 0.06~degrees ($\sim$720~km at the disk center).

The magnetic field modeling involves two computational tasks. First, a preprocessing of the observed vector magnetic field data \citep{2006SoPh..233..215W} is required in order to obtain a force-free consistent bottom boundary. Second, based on the preprocessed boundary data, the NLFF corona above the AR is approximated by applying a combination of the improved optimization scheme of \cite{2010A&A...516A.107W} and a multi-scale approach \citep{2008SoPh..247..249W}.

In a recent study, \cite{2020A&A...643A.153T} investigated the effect of particular choices of free model parameters, both during preprocessing and optimization, onto subsequent magnetic helicity computation. It was found in their study that the application of preprocessing with default parameters (including the smoothing of the originally observed vector data) improves the NLFF model quality, in terms of force-freeness and solenoidality \citep[see also][]{2006SoPh..233..215W}. It was also shown that further improvement of the solenoidality of the NLFF models can be achieved by enhancing the relative importance of the volume-integrated divergence compared to the importance of the volume-integrated Lorentz force, minimized during optimization. Therefore, in our study, we analyze two time series of NLFF models for each AR, one time series of equally strong weighting of the 3D force- and divergence freeness \citep[$w_f=w_d=1$; see Eq. (4) in][]{2010A&A...516A.107W}, and one time series with an enhanced weighting of the volume-integrated divergence (on the expense of force-freeness; $w_f=1$, $w_d=2$).

To determine the force-free consistency of the NLFF models, we calculate the current-weighted angle between the modeled magnetic field and the electric current density, $\thetaj$ \citep[e.g.,][]{2006SoPh..235..161S}. For a completely force-free field, $\thetaj=0$. We note here that the average $\thetaj$ in our NLFF models, for all of the considered ARs and most times during the analysis time intervals, is $\lesssim15^\circ$ (see \href{CW_theta}{\gm{Fig.}~(\ref{CW_theta})} in the \href{appendixA}{Appendix~\ref{appendixA}}). In order to determine the degree of solenoidality of a given $\Bvec$, \cite{2013A&A...553A..38V} introduced a relative measure which expresses the fraction of the total magnetic energy ($\Etot$) that is related to the non-zero divergence ($\Ediv$), in the form $\Edivprime$. For completeness, we show the corresponding values for each AR in \href{Ediv_E}{\gm{Fig.}~(\ref{Ediv_E})} in the \href{appendixA}{Appendix~\ref{appendixA}}. Dedicated studies showed that a maximum of $\Edivprime=0.1$ is to be tolerated for subsequent magnetic helicity computation \citep{2016SSRv..201..147V,2019ApJ...880L...6T,2019A&A...628A..50M}. Correspondingly, we analyze the coronal magnetic helicity and energy (and related quantities) only for those NLFF models that satisfy the above criterion, i.e., omit the analysis of those models for which $\Edivprime>0.1$.

In order to be able to compute the relative helicities from \href{eq:hv}{\gm{Eqs.}~(\ref{eq:hv})}--\href{eq:hv}{(\ref{eq:hpj})}, we apply the method of \cite{2011SoPh..272..243T} to compute the vector potentials $\Avec$ and $\Apvec$, corresponding to $\Bvec$ and $\Bpvec$, respectively. The method solves systems of partial differential equations to obtain the vector potentials $\Avec$ and $\Apvec$, using the Coulomb gauge, $\nabla\cdot\Avec=\nabla\cdot\Apvec=0$. The method has been tested in \cite{2016SSRv..201..147V} and is known to deliver helicities in line with that of other methods.
 
\subsection{Quantities investigated}
For each time instance of the NLFF model time series of the ARs, we compute the total ($\Etot$), potential ($\Epot$) and free ($\Efree$) magnetic energy, as well as the total ($\hv$), volume-threading ($\hpj$) and current-carrying ($\hj$) relative helicities. We compute averages of those quantities, together with a corresponding spread, for those time instances where both NLFF models qualify for analysis, i.e., where $\Edivprime\leq0.1$ for both. If only one of the NLFF models at a given time instant qualifies for analysis, we keep only the corresponding single values.
We note here that in addition we excluded one data point from the time series modeling of AR 12673 due to an unusual high value of $\mathbf{\theta_j>35^\circ}$ at 11:35~UT (see \href{CW_theta}{\gm{Fig.}~\ref{CW_theta}(a)}).
Subsequently, we calculate the intensive quantities such as helicity ratio, $\hjprime$, energy ratio, $\Efprime$, and the normalized helicities, $\hvnorm$ and $\hjnorm$. Here, the magnetic flux used for normalization is calculated as half of the total unsigned flux that is shown in \href{goes_absF}{\gm{Fig.}~(\ref{goes_absF})}, i.e., from the relation $\phi^{\prime}=\frac{1}{2}\int_{S(z=0)}^{}|{B_{z}}|{\rm ~d}S$. 

From the time profiles of the quantities introduced above, we estimate a "characteristic" pre-flare level as follows. Taking into account all data points within a time window of five hours prior to flare onset, we compute the mean value (hereafter called "time-averaged" and denoted by angular brackets) and standard deviation. For completeness we note that the retrieved values do not significantly vary if the width of the pre-flare time windows is between 3-6 hours. Only for AR 11166, the time-averaged values are significantly different if the pre-flare time window is less than 3 hours.

\section{Results}
In the following \href{ss:Extensive}{\gm{Sect.}~\ref{ss:Extensive}}, we describe the evolution of the extensive quantities, including the unsigned magnetic flux, $\flux$, helicities, $\hv$, $\hpj$, $\hj$, and the magnetic energies, $\Etot$ and $\Efree$. In \href{ss:Intensive}{\gm{Sect.}~\ref{ss:Intensive}}, we discuss the evolution of intensive quantities derived from the extensive quantities. To further explore these quantities in detail, we analyze the characteristic time-averaged pre-flare values of the extensive and intensive quantities in \href{ss:preflare}{\gm{Sect.}~\ref{ss:preflare}}.
\subsection{Extensive Quantities (Magnetic flux, magnetic energies and relative helicities)}
\label{ss:Extensive}

\begin{figure*}[htp]
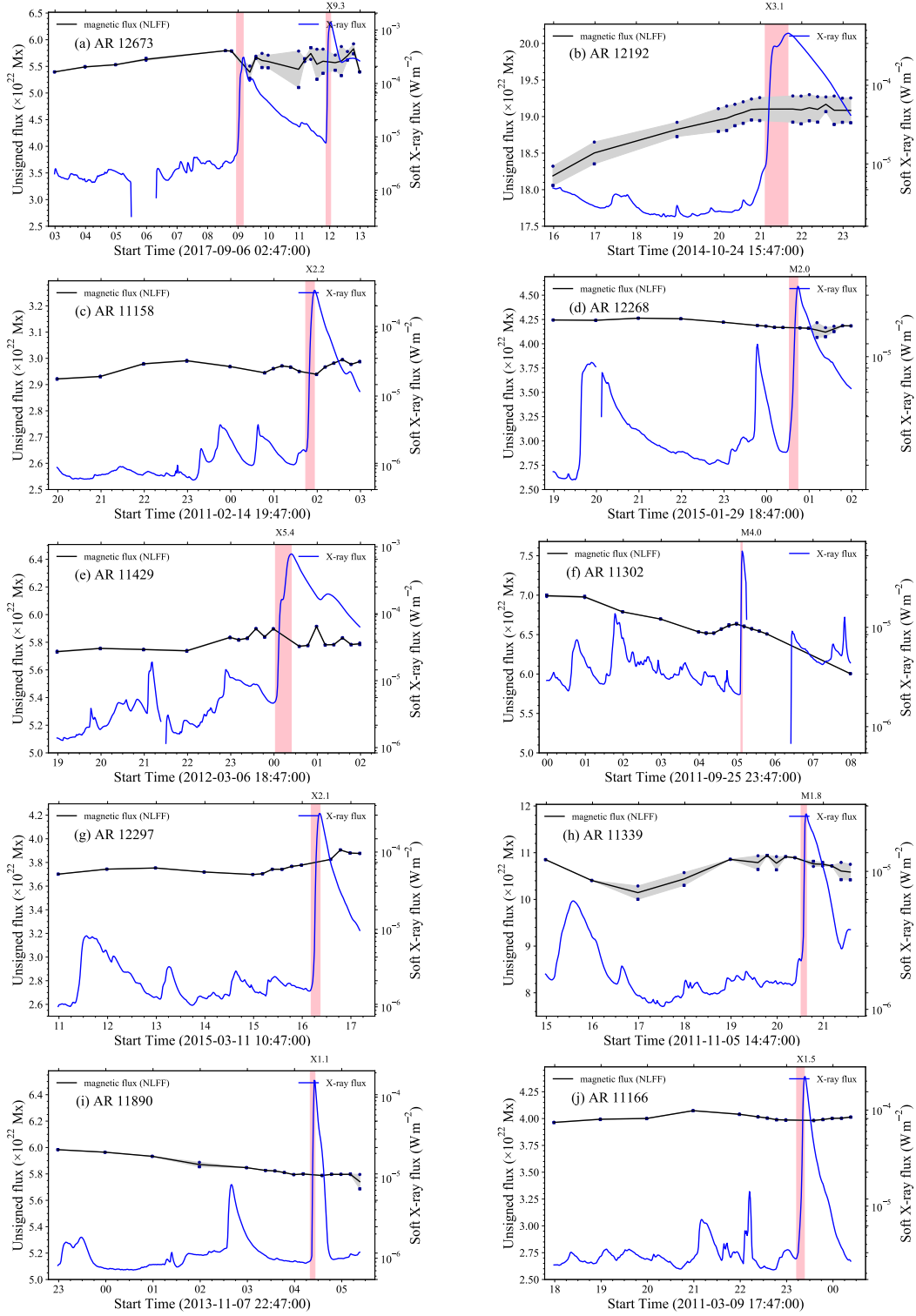

    \captionsetup{width=\linewidth}
    \jktfig{goes_absF}{.pdf}
    \caption{\gm{Time evolution of the magnetic flux, $\flux$, and the GOES soft X-ray flux for the 10 ARs under study. Quantities for ARs productive of large eruptive and confined flares are shown in the left and right column, respectively.} Black curve represents the average of the magnetic fluxes calculated from NLFF bottom boundary fields based on enhanced (standard) divergence freeness as indicated by squares (bullets). The gray shaded areas mark the spread of $\flux$. Blue curve shows the evolution of GOES soft-X ray flux. Vertical bar indicate the impulsive flare phase.}
    \label{goes_absF}
\end{figure*}

\begin{figure*}[htp]
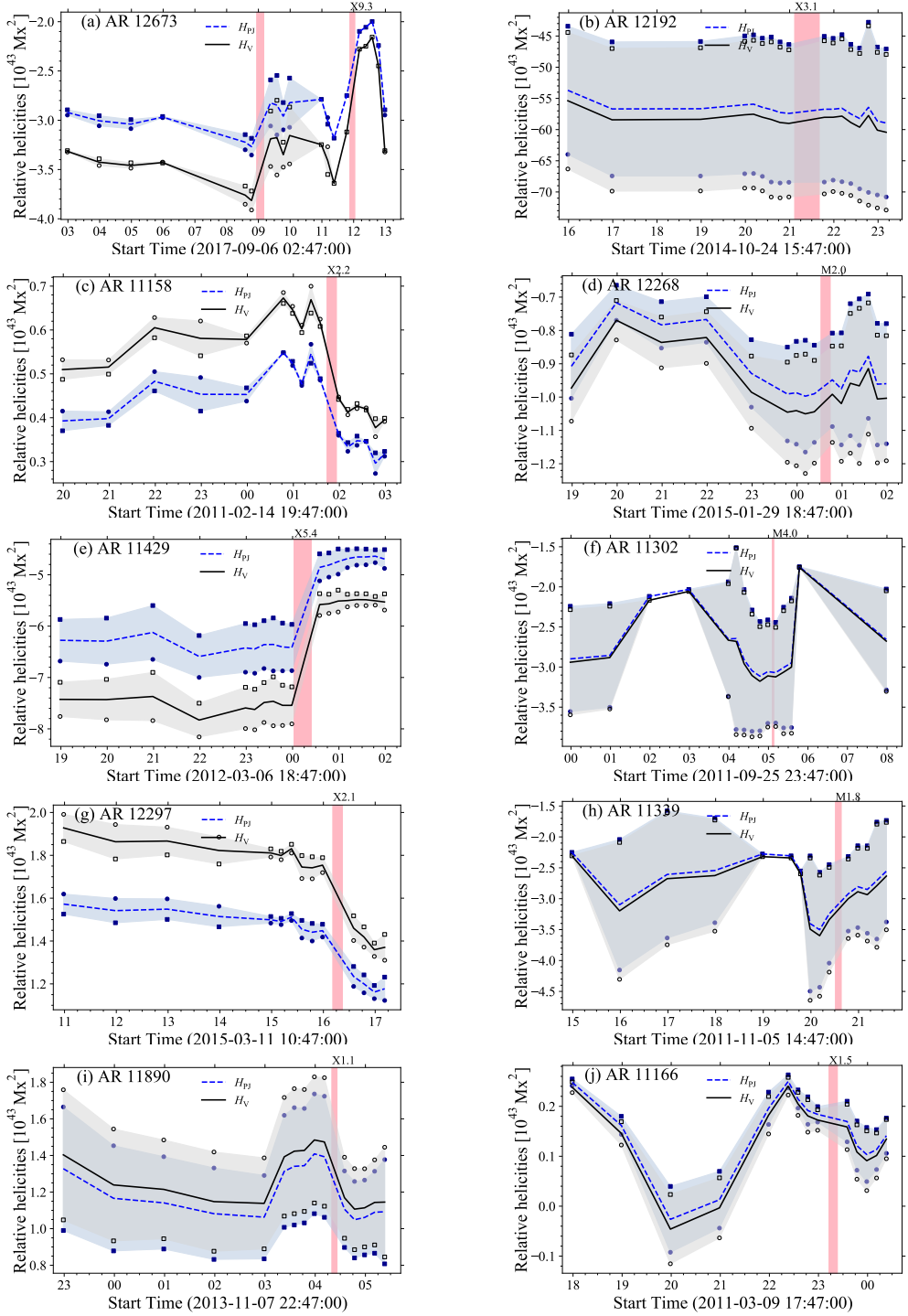

    \captionsetup{width=\linewidth}
    \fig{H_Hpj}{.pdf}
    \caption{Time evolution of \gm{the relative} magnetic helicity $\hv$ and volume threading helicity $\hpj$ for the 10 ARs under study. \gm{Quantities for ARs productive of large eruptive and confined flares are shown in the left and right column, respectively.} Squares (bullets) indicate the solutions based on enhanced (standard) divergence freeness. The gray (light-blue) shaded areas mark the spread of $\hv$ ($\hpj$), and the solid-black (dashed-blue) curve indicate the respective average values of $\hv$ ($\hpj$). The vertical bar marks the impulsive flare phase.}
    \label{H_Hpj}
\end{figure*}

\begin{figure*}[htp]
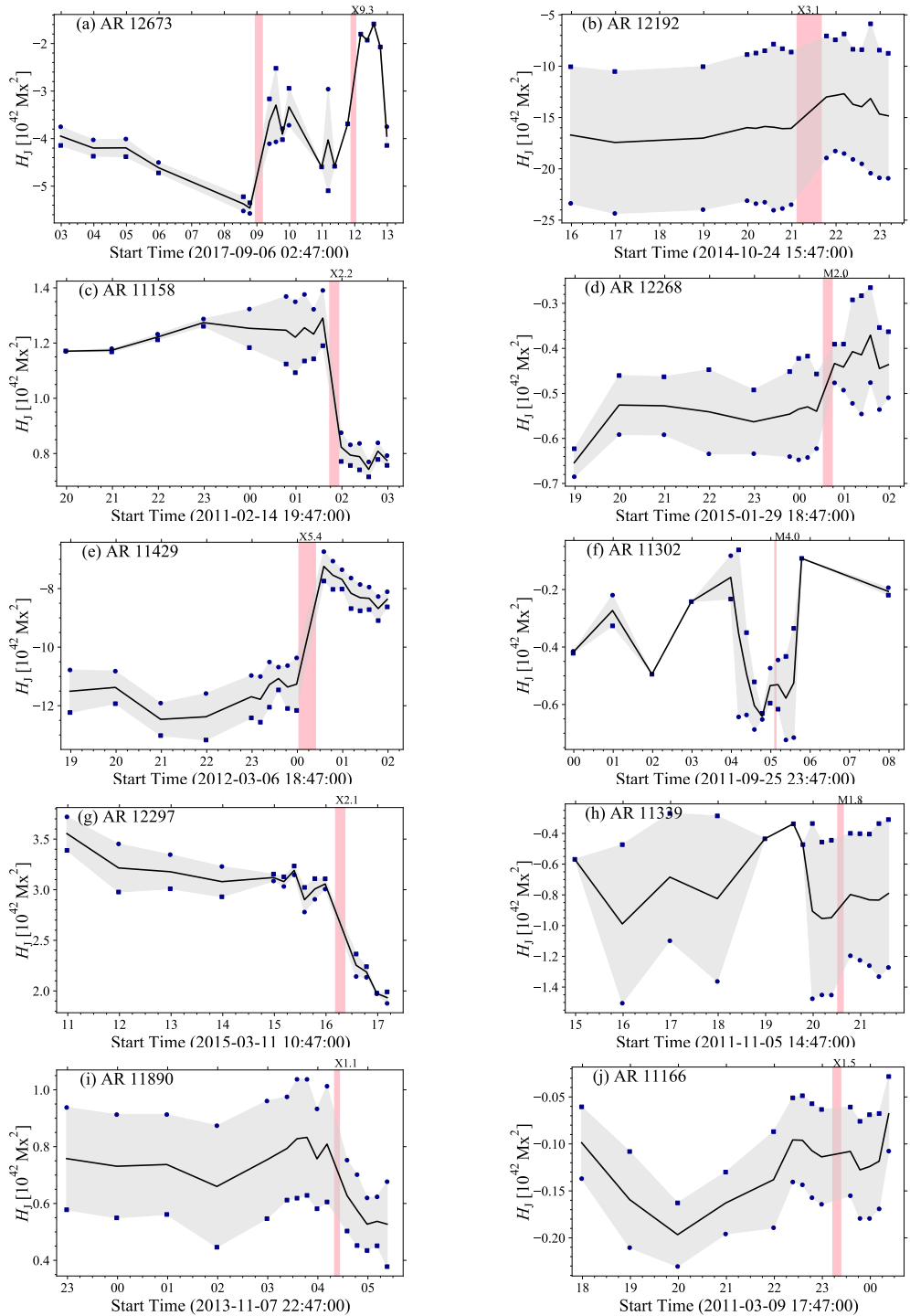

    \captionsetup{width=\linewidth}
    \fig{Hj}{.pdf}
    \caption{Time evolution of \gm{the current-carrying} helicity $\hj$ for the 10 ARs under study. \gm{Quantities for ARs productive of large eruptive and confined flares are shown in the left and right column, respectively.} Squares (bullets) indicate the solutions based on enhanced (standard) divergence freeness. The gray shaded areas mark the spread of $\hj$, and the black curve indicate the average values. The vertical bar marks the impulsive flare phase.}
    \label{Hj}
\end{figure*}

\begin{figure*}[htp]
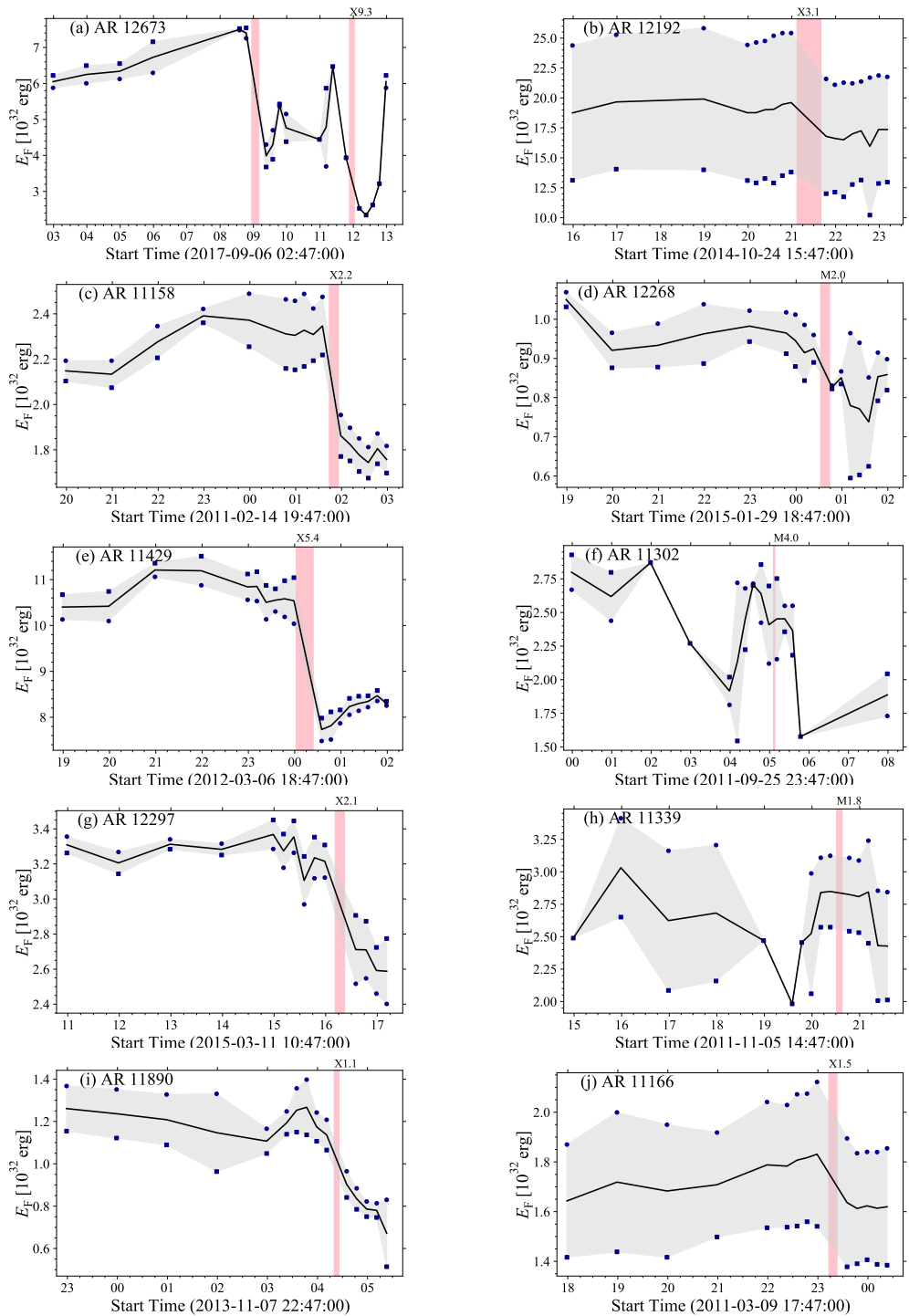

    \captionsetup{width=\linewidth}
    \fig{Ef}{.pdf}
    \caption{Time evolution of \gm{the free} magnetic energy $\Efree$ for the 10 ARs under study. Layout as in Fig.~\ref{Hj}}
    \label{Ef}
\end{figure*}

\begin{figure*}[htp]
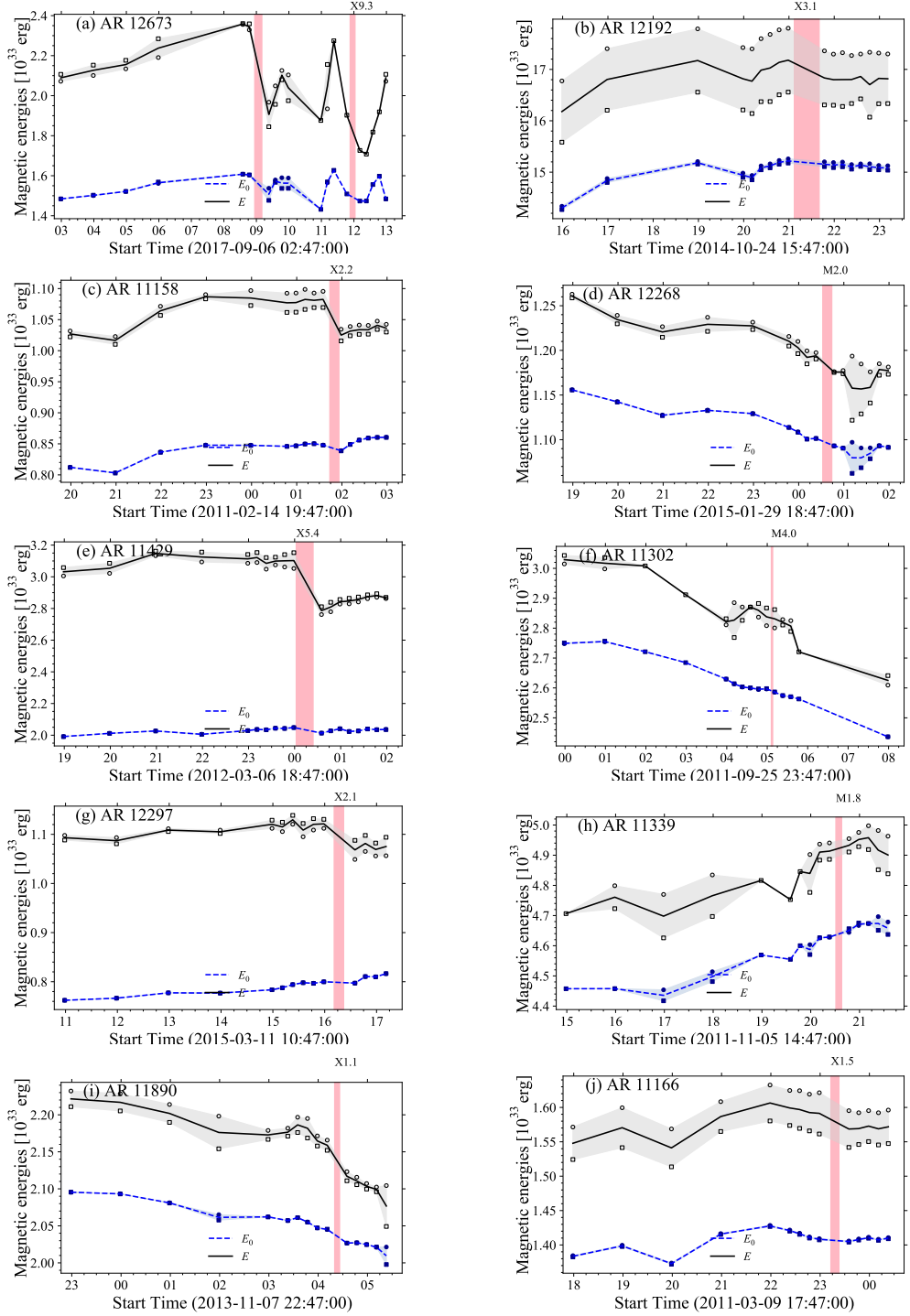

    \captionsetup{width=\linewidth}
    \fig{E_Ep}{.pdf}
    \caption{Time evolution of \gm{the total} magnetic energy $\Etot$ and energy of potential magnetic field $\Epot$ for the 10 ARs under study. Layout as in Fig.~\ref{H_Hpj}.}
    \label{E_Ep}
\end{figure*}

\begin{figure*}[htp]
    \captionsetup{width=\linewidth}
    \fig{H_J_H}{.pdf}
    \caption{Time evolution of \gm{the helicity} ratio $\hjprime$ for the 10 ARs under study. Layout as in Fig.~\ref{Hj}}
    \label{H_J_H}
\end{figure*}

\begin{figure*}[htp]
    \captionsetup{width=\linewidth}
    \fig{Ef_E}{.pdf}
    \caption{Time evolution of \gm{the energy} ratio $\Efprime$ for the 10 ARs under study. Layout as in Fig.~\ref{Hj}}
    \label{Ef_E}
\end{figure*}

\begin{figure*}[htp]
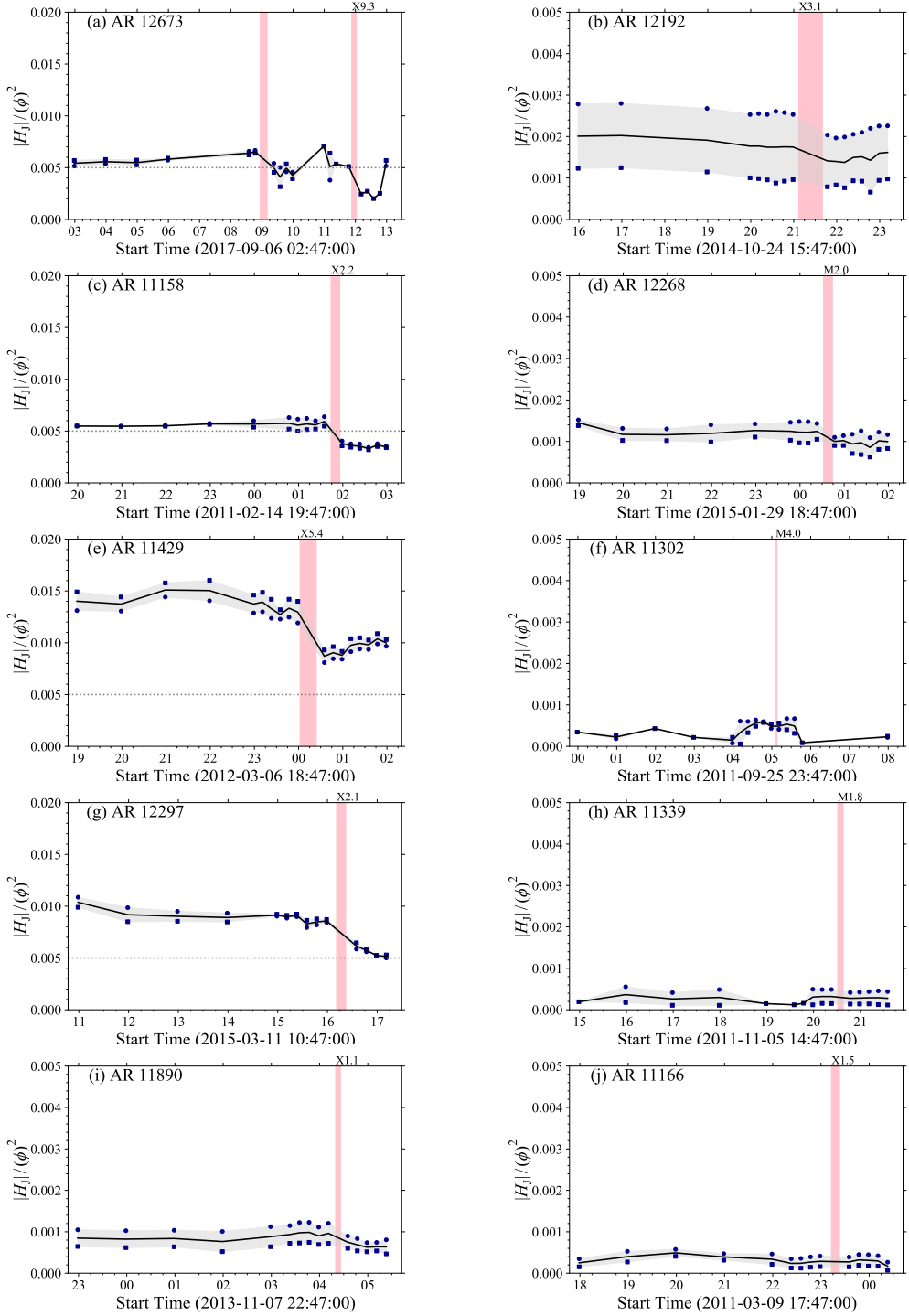

    \captionsetup{width=\linewidth}
    \fig{Hj_0_5_phi2}{.pdf}
    \caption{Time evolution of \gm{the normalized} current-carrying helicity $\hjnorm$ for the 10 ARs under study. Layout as in Fig.~\ref{Hj}}
    \label{Hj_0_5_phi2}
\end{figure*}

\begin{figure*}[htp]
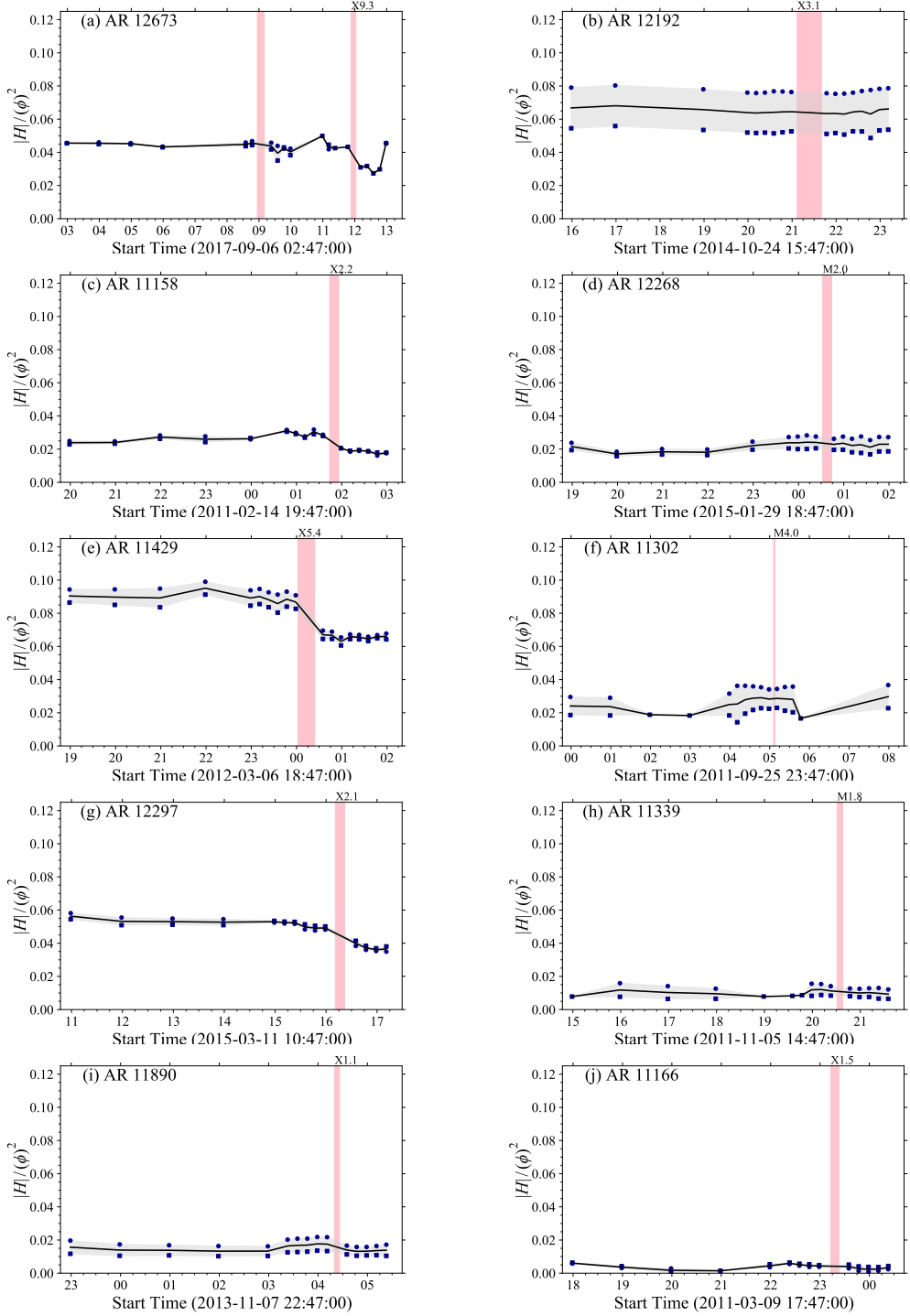

    \captionsetup{width=\linewidth}
    \fig{H_0_5_phi2}{.pdf}
    \caption{Time evolution of \gm{the normalized} relative magnetic helicity $\hvnorm$ for the 10 ARs under study. Layout as in Fig.~\ref{Hj}}
    \label{H_0_5_phi2}
\end{figure*}

For each of the ARs under study, the time evolution of the magnetic field related quantities is described in detail, and discussed in context with previous studies (if existing). For all quantities we compute average values from the two different NLFF model time series at each time instant. The corresponding spread (except for time instances where  only one valid NLFF solution is found) is indicated by an adjacent shaded area.

\href{goes_absF}{Figure~\ref{goes_absF}} shows the evolution of the unsigned magnetic flux (black curve) along with the GOES soft X-ray light curves (blue) for each of the 10 ARs. 
The unsigned magnetic flux values range over about one order of magnitude (from $2.9\times10^{22}\, \si{Mx}$ to $1.9\times10^{23}\, \si{Mx}$).
\href{H_Hpj}{Figure~\ref{H_Hpj} shows the time evolution of the magnetic helicity, $\hv$ (black curve), and the contribution of volume threading helicity, $\hpj$ (blue curve). The time evolution of the non-potential helicity, $\hj$, for each AR, is
shown in \href{Hj}{\gm{Fig.}~\ref{Hj}}. 
We find the absolute values of the relative helicities, $\hv$, $\hpj$ and $\hj$, to span roughly three orders of magnitude, in the approximate range of $3.3\times10^{40} \, \si{{Mx}^{2}}$ to $6\times10^{44} \, \si{{Mx}^{2}}$.} 
\href{Ef}{Figure~\ref{Ef} displays the evolution of the free magnetic energy, $\Efree$. \href{E_Ep}{Figure~\ref{E_Ep}} shows the evolution of total magnetic energy $\Etot$ (black curve) and the energy of the potential magnetic fields $\Epot$ (blue curve). ARs hosting a higher unsigned magnetic flux also harbor a higher $\Etot$.
In contrast to total energy $\Etot$, no clear relation to the underlying unsigned magnetic flux is obtained for the free magnetic energy $\Efree$, which may be expected, since it is related to the presence of electric currents.}
In the following, we discuss the time evolution of the average magnetic helicities and energies for each event:

\begin{itemize}
\item For AR 12673, the relative magnetic helicity, $\hv$, was negative with magnitudes up to $\approx3.8 \times 10^{43} \, \si{{Mx}^{2}}$ prior to the onset of the first X-class flare (black curve in \href{H_Hpj}{\gm{Fig.}~\ref{H_Hpj}(a)}). Similar values have been reported by \cite{2019A&A...628A..50M}, \cite{2019ApJ...872..182V}, \cite{2019A&A...628A.114P}, and \cite{2020A&A...643A.153T}. Individual contributions of $\hj$ (\href{Hj}{\gm{Fig.}~\ref{Hj}(a)}) and $\hpj$ (blue line in \href{H_Hpj}{\gm{Fig.}~\ref{H_Hpj}(a)}) display the same sign and temporal evolution as $\hv$. Also in agreement to these earlier studies, we find maximum pre-flare values of the free magnetic energy, $\Efree\approx7.5\times10^{32}$ erg, before the onset of the eruptive flare (\href{Ef}{\gm{Fig.}~\ref{Ef}(a)}).
\item The relative magnetic helicity, $\hv$, in AR 11158 before the start of the flare attains a maximum value of $\approx6.7\times10^{42} \, \si{{Mx}^{2}}$ (\href{H_Hpj}{\gm{Fig.}~\ref{H_Hpj}(c)}). The time profile and magnitudes of the relative helicity are in overall agreement with the results presented by \cite{2012ApJ...752L...9J,2015RAA....15.1537J} and \cite{2019ApJ...887...64T}. Helicities, $\hj$ and $\hpj$, as shown in \href{Hj}{\gm{Figs.}~\ref{Hj}(c)} and\href{H_Hpj}{~\ref{H_Hpj}(c)} respectively, are positive. The free magnetic energy is $\Efree\approx 2.3\times 10^{32}$ erg (\href{Ef}{\gm{Fig.}~\ref{Ef}(c)}) for two hours before the flare onset. Similar values were reported by \cite{2012ApJ...748...77S} and \cite{2019ApJ...887...64T}.
\item For AR 11429, we find $\hv$ to be negative and on the order of $\approx 7.5 \times 10^{43}\, \si{{Mx}^{2}}$ before the flare (\href{H_Hpj}{\gm{Fig.}~\ref{H_Hpj}(e)}), in accordance to the sign of helicity obtained, \gm{for example}, by \cite{2018AdSpR..61..673Y} and \cite{2014SoPh..289.2957E}. $\hj$ and $\hpj$ are also negative, and show a similar trend as $\hv$ (see \href{Hj}{\gm{Figs.}~\ref{Hj}(e)} and\href{H_Hpj}{~\ref{H_Hpj}(e)}, respectively). The free magnetic energy is considerably high with pre-flare $\Efree\approx 1.1\times 10^{33}$ erg (\href{Ef}{\gm{Fig.}~\ref{Ef}(e)}).
\item In AR 12297, the magnetic helicities, $\hv$, $\hpj$, $\hj$, are with positive sign, and gradually decrease throughout the considered time period (see \href{H_Hpj}{\gm{Figs.}~\ref{H_Hpj}(g)} and\href{Hj}{~\ref{Hj}(g), respectively}). $\hv$ prior to the flare onset is $\approx 1.80 \times 10^{43}\, \si{{Mx}^{2}}$. The free magnetic energy ($\Efree$) is almost constant ($\Efree\approx3.3\times10^{32}$ erg) prior to the flare onset (\href{Ef}{\gm{Fig.}~\ref{Ef}(g)}).
\item In AR 11890, $\hv$ is positive and increases up to $\approx1.5\times10^{43}\, \si{{Mx}^{2}}$ before the flare (\href{H_Hpj}{\gm{Fig.}~\ref{H_Hpj}(i)}). $\hj$ and $\hpj$ are also positive, and show a similar trend as $\hv$ (see \href{Hj}{\gm{Figs.}~\ref{Hj}(i)} and\href{H_Hpj}{~\ref{H_Hpj}(i)}, respectively). The free energy, $\Efree$, is $\approx1.2\times10^{32}$ erg prior to the flare onset (\href{Ef}{\gm{Fig.}~\ref{Ef}(i)}). Both, $\hj$ and $\Efree$ are lowest among all of the studied CME-productive ARs.
\item Magnetic helicities in AR 12192 around the confined X3.1 flare are negative in sign (\href{H_Hpj}{\gm{Figs.}~\ref{H_Hpj}(b)} and\href{Hj}{~\ref{Hj}(b)}). The spread in helicity and energy related quantities is higher in AR 12192 in comparison to other ARs. This is because the spread in the unsigned flux is approximately 10 times higher than the other ARs. Consequently, this spread in the underlying model flux values also causes a higher spread in rest of the model based quantities \gm{such as} magnetic energy and helicity. The absolute value of the relative magnetic helicity, $\hv\approx58\times10^{43}\, \si{{Mx}^{2}}$ is more or less constant through out the time series. The free energy, $\Efree$ %(last column of \textbf{\href{t2}{Table~\ref{t2}}} and \href{Ef}{\gm{Fig.}~\ref{Ef}(b)})
(\href{Ef}{\gm{Fig.}~\ref{Ef}(b)}), is $\approx1.9\times10^{33}$ erg before the flare. Our values are in line with those of the earlier study of \cite{2019ApJ...887...64T}. We note that \cite{2015RAA....15.1537J} reported corresponding estimates of $\hv$ and $\Efree$ by a factor of 10 lower. 
\item Around the confined M2.0 flare, the relative magnetic helicity of AR 12268 is negative, and $\hv$ increases up to a peak value of $\approx -1 \times 10^{43}\, \si{{Mx}^{2}}$ before the flare (see \href{H_Hpj}{\gm{Fig.}~\ref{H_Hpj}(d)}). 
$\hj$ is nearly constant, while $\hpj$ is increasing, before the flare (\href{Hj}{\gm{Fig.}~\ref{Hj}(d)} and \href{H_Hpj}{~\ref{H_Hpj}(d)}, respectively). 
The pre-flare values of the free energy, $\Efree \approx 0.9\times10^{32}$ erg, are the lowest among all the ARs studied here.
\item Around the confined M4.0 flare produced by AR 11302, $\hv$ is negative and the absolute values of $\hv$ ranges between $-1.75\times10^{43}\, \si{{Mx}^{2}}$ and $-3.2\times10^{43}\, \si{{Mx}^{2}}$ during the analyzed time interval (see \href{H_Hpj}{\gm{Fig.}~\ref{H_Hpj}(f)}).
The free energy, $\Efree$, fluctuates between $1.5\times10^{32}$ erg and $3\times10^{32}$ erg (\href{Ef}{\gm{Fig.}~\ref{Ef}(f)}).
\item Around the confined M1.8 flare in AR 11339, $\hv$ is negative and the values are in the range $(-2,-3.6)\times10^{43}\, \si{{Mx}^{2}}$ (\href{H_Hpj}{\gm{Fig.}~\ref{H_Hpj}(h)}). The decomposition of the relative magnetic helicity, $\hj$ and $\hpj$ are also negative and largely vary throughout the time series (\href{Hj}{\gm{Fig.}~\ref{Hj}(h)} and \href{H_Hpj}{~\ref{H_Hpj}(h)}). Unlike other ARs in this study, the total magnetic energy, $\Etot$, is increasing during the flare (\href{E_Ep}{\gm{Fig.}~\ref{E_Ep}(h)}), and the free energy, $\Efree$, is almost constant ($\Efree \approx 2.8\times10^{32}$ erg) at the time of the flare (\href{Ef}{\gm{Fig.}~\ref{Ef}(h)}).
\item Around the confined X1.5 flare produced by AR 11166, $\hj$ and $\hpj$ are of opposite sign throughout the time series, except for the short interval when $\hpj$ reverses from positive to marginally negative and then back to positive values (see \href{Hj}{\gm{Fig.}~\ref{Hj}(j)} \&\href{H_Hpj}{~\ref{H_Hpj}(j)}). Also, the average relative magnetic helicity decreases from  a value of $2.4 \times 10^{42}\, \si{{Mx}^{2}}$ at 18:00 UT to $-0.4 \times 10^{42}\, \si{{Mx}^{2}}$ at 20:00 UT (\href{H_Hpj}{\gm{Fig.}~\ref{H_Hpj}(j)}). \cite{2012ApJ...761...86V} in their study of AR 11166 observed patches of negative helicity flux embedded in the positive helicity flux site of the flare, with similar values of magnetic helicity accumulation as in our case. As the total relative helicity decreases from a positive to negative value, $\hj$ does not change significantly. The free energy increases to a maximum value of $\Efree\approx1.8\times10^{32}$ erg prior to the onset of the flare. 
\end{itemize}

In summary, all our overall helicity budgets agree in sign and magnitude with that recovered in previous studies (if existing). $\hv$ is negative in six ARs (AR 12673, AR 12192, AR 12268, AR 11302, AR 11339), and positive in three ARs (AR 11158, AR 12297, AR 11890). Only AR 11166 exhibits a reversal of the sign of helicity, i.e., a transition from a positive to a marginally negative coronal helicity budget. The absolute values of $\hv$ decrease considerably during the eruptive flares. For the majority of ARs, $\hj$ shares the sign of $\hv$. Only for AR 11166, we find the sign of $\hj$ being opposite to that of $\hv$ during most of the time interval studied. For all the analyzed ARs, the absolute values of $\hj$ decrease largely during the eruptive flares. $\hj$ values are not distinctive to the magnitude or type of the flares hosted by the individual ARs. In all of the analyzed ARs, $\hpj$ shares the sign of $\hv$. Similar to $\hj$, we see a notable decrease in $\hpj$ during the eruptive flares. The absolute values of $\hv$, $\hj$ and $\hpj$ span a broad range, and appear not to be related to the size (magnitude) or type (confined or eruptive) of the observed flares.

Similarly, the values of the total magnetic energy $\Etot$ of the ARs are not distinguishable for different magnitudes or type of the flares. In general, $\Etot$ shows a decreasing trend during the flares, except in case of AR 11339, where $\Etot$ is increasing during the confined M4.0 flare. The free magnetic energy, $\Efree$, in general, also decreases during the flares. Only for AR 11339, $\Efree$ remains almost constant over the course of the flare. Also, similar to other extensive quantities, $\Efree$ does not show any discrimination to the magnitude or type of flares produced.

\subsection{Intensive quantities (non-potential helicity ratio, free energy ratio, and normalized helicities)}\label{ss:Intensive}

Based on the time evolution of the extensive quantities presented in \href{ss:Intensive}{\gm{Sect.}~\ref{ss:Extensive}}, we are able to study the corresponding time evolution of the intensive ones. In the following, overall trends visible in the deduced time profiles of the intensive quantities are qualitatively described, while a quantitative analysis is presented in \href{ss:preflare}{\gm{Sect.}.~\ref{ss:preflare}}. In \gm{Figs.}~\href{H_J_H}{\ref{H_J_H}}--\href{H_0_5_phi2}{\ref{H_0_5_phi2}}, the time evolution of helicity and energy-based intensive quantities are shown, including the non-potential helicity ratio ($\hjprime$), the free energy ratio ($\Efprime$), the normalized non-potential helicity ($\hjnorm$), and the normalized total helicity ($\hvnorm$). 

\href{H_J_H}{Figure~\ref{H_J_H}} shows the time evolution of the non-potential helicity ratio, $\hjprime$. Importantly, we find that in general the level of $\hjprime$ before the analyzed flare is significantly higher when it is associated to a CME. Furthermore, $\hjprime$ shows a decreasing trend during all the eruptive flares (see left column in \href{H_J_H}{\gm{Fig.}~\ref{H_J_H}}). Similarly, the pre-flare values of $\Efprime$, as shown in \href{Ef_E}{\gm{Fig.}~\ref{Ef_E}}, are significantly larger for the eruptive flares (see left column in \href{Ef_E}{\gm{Fig.}~\ref{Ef_E}}). 

The time evolution of the normalized non-potential helicity, $\hjnorm$, is shown in \href{Hj_0_5_phi2}{\gm{Fig.}~\ref{Hj_0_5_phi2}} and also exhibits a similar behavior as the non-potential helicity and free energy ratios. In most of the eruptive flares, the levels of $\hjnorm$ are elevated before the flare and significantly decrease over the course of the flare (see left column in \href{Hj_0_5_phi2}{\gm{Fig.}~\ref{Hj_0_5_phi2}}).

The time evolution of the normalized total helicity, $\hvnorm$ shown in \href{H_0_5_phi2}{\gm{Fig.}~\ref{H_0_5_phi2}}, does not reveal characteristically different pre-flare levels for confined and eruptive flares. Its flare-related change, however, seems to be more pronounced for eruptive flares (see left column in \href{H_0_5_phi2}{\gm{Fig.}~\ref{H_0_5_phi2}}).

Given the above trends, only AR 11890 appears to represent an exceptional case, as it exhibits values of intensive quantities before the eruptive flare that are otherwise typical for confined flares. Furthermore, though AR 11166 exhibits immediate pre-flare values of the intensive quantities similar to that of other confined flares, we still notice an elevation of the non-potential helicity ratio, $\hjprime$, during three hours well before the flare onset. Possible explanations for these exceptions are discussed in \href{Discussion}{\gm{Sect.}~\ref{Discussion}}.

\begin{table*}[ht]
%\captionabove
\caption{\label{t2} NOAA number, and time-averaged pre-flare values of the extensive quantities for the ten ARs under study. Time averages are computed from all data points within a five-hour time window prior to the nominal start time of the flare. The top five rows refer to eruptive flares, whereas the bottom five rows refer to confined flares.}
\centering
\begin{tabular}{lcccccc}
\hline\hline
%rule{0pt} to create space after horizontal line
\rule{0pt}{4ex} NOAA & \makecell{$\fluxmean$\\($10^{22}\; \si{Mx}$)} & \makecell{$\hvmean$\\($10^{43}\; \si{{Mx}^{2}}$)} & \makecell{$\hjmean$\\($10^{42}\; \si{{Mx}^{2}}$)} & \makecell{$\hpjmean$\\($10^{43}\; \si{{Mx}^{2}}$)} & \makecell{$\Etotmean$\\($10^{33}\; \si{erg}$)} & \makecell{$\Efreemean$\\($10^{32}\; \si{erg}$)}\\
\hline
 12673 &$5.62\pm0.12$ &$-3.39\pm0.25$ &$-4.19\pm0.74$ &$-2.97\pm0.18$ &$2.09\pm0.17$ &$5.30\pm1.29$\\ %13 \& 9 min 
 11158 &$2.96\pm0.02$ &$0.61\pm0.05$ &$1.24\pm0.03$ &$0.49\pm0.05$ &$1.07\pm0.02$ &$2.31\pm0.07$ \\%12 min
 11429 &$5.82\pm0.06$ &$-7.54\pm0.13$ &$-11.63\pm0.47$ &$-6.38\pm0.12$ &$3.11\pm0.03$ &$10.74\pm0.28$ \\%22 min
 12297 &$3.74\pm0.03$ &$1.80\pm0.05$ &$3.09\pm0.09$ &$1.49\pm0.04$ &$1.11\pm0.01$ &$3.26\pm0.08$ \\%11 min
 11890 &$5.85\pm0.06$ &$1.33\pm0.13$ &$0.77\pm0.05$ &$1.25\pm0.13$ &$2.18\pm0.02$ &$1.19\pm0.05$ \\
\hline
 12192 &$18.94\pm0.19$ &$-58.28\pm0.5$ &$-16.31\pm0.54$  &$-56.65\pm0.49$ &$16.99\pm0.16$ &$19.28\pm0.41$ \\%33 min
 12268 &$4.21\pm0.04$ &$-0.95\pm0.11$ &$-0.54\pm0.01$ &$-0.90\pm0.11$ &$1.21\pm0.02$ &$0.94\pm0.02$ \\%12 \& 7 min
 11302 &$6.65\pm0.14$ &$-2.76\pm0.38$ &$-0.42\pm0.16$ &$-2.71\pm0.38$ &$2.89\pm0.07$ &$2.45\pm0.28$ \\%2 min
 11339 &$10.68\pm0.27$ &$-2.91\pm0.47$ &$-0.73\pm0.24$ &$-2.84\pm0.45$ &$4.81\pm0.07$ &$2.61\pm0.29$ \\%7 min
 11166 &$4.01\pm0.03$ &$0.13\pm0.10$ &$-0.13\pm0.03$ &$0.15\pm0.09$ &$1.59\pm0.02$ &$1.77\pm0.05$ \\
\hline
\end{tabular}
\end{table*}

\begin{table*}[ht]
\caption{\label{t3} NOAA number, GOES SXR flare class, and time-averaged pre-flare values of the intensive quantities for the ten ARs under study. Time averages are computed from all data points within a five-hour time window prior to the nominal flare start time. The top five rows refer to eruptive flares, whereas the bottom five rows refer to confined flares.}
\centering
\begin{tabular}{lccccccc}
\hline\hline
\rule{0pt}{4ex}NOAA & SXR class & $\hjprimemean$ &$\Efprimemean$ &\makecell{$\hvnormmean$\\$(10^{-2})$} &\makecell{$\hjnormmean$\\$(10^{-2})$}\\
\hline
 12673 &X9.3& $0.12\pm0.01$& $0.25\pm0.04$& $4.35\pm0.27$& $0.54\pm0.09$\\ %13 \& 9 min 
 11158 &X2.2& $0.20\pm0.02$& $0.22\pm0.00$
 &$2.78\pm0.21$& $0.57\pm0.01$\\%12 min
 11429 &X5.4& $0.16\pm0.01$& $0.35\pm0.01$& $8.92\pm0.25$ & $1.38\pm0.08$\\%22 min
 12297 &X2.1& $0.17\pm0.00$& $0.29\pm0.01$
 &$5.16\pm0.16$ & $0.89\pm0.03$\\%11 min
 11890 &X1.1& $0.06\pm0.00$& $0.05\pm0.00$
 &$1.55\pm0.18$ & $0.09\pm0.01$\\
\hline
 12192 &X3.1& $0.03\pm0.00$& $0.11\pm0.00$ 
 &$6.48\pm0.14$& $0.18\pm0.01$\\%33 min
 12268 &M2.0& $0.06\pm0.01$& $0.08\pm0.00$
 &$2.14\pm0.29$ & $0.12\pm0.00$\\%12 \& 7 min
 11302 &M4.0& $0.02\pm0.01$& $0.08\pm0.01$
 &$2.50\pm0.39$ & $0.04\pm0.02$\\%2 min
 11339 &M1.8& $0.02\pm0.00$& $0.05\pm0.01$
 &$1.02\pm0.16$ & $0.03\pm0.01$\\%7 min
 11166 &X1.5& $0.14\pm0.13$& $0.11\pm0.00$
 &$0.39\pm0.15$ & $0.03\pm0.01$\\
\hline
\end{tabular}
\end{table*}

\subsection{Characteristic pre-flare levels}\label{ss:preflare}

In the following, we analyze the overall properties of time-averaged pre-flare levels of \gm{all of the quantities.} \href{t2}{Table~\ref{t2}} lists the time-averaged pre-flare levels of the extensive quantities ($\hv$, $\hj$, $\hpj$, $\Etot$, $\Efree$)  for our sample of ARs, while in Table 3, we list the time-averaged pre-flare levels of the intensive quantities ($\hjprime$, $\Efprime$, $\hvnorm$, $\hjnorm$).

\gm{Summarizing the time-averaged pre-flare levels of the quantities,} we find absolute time-averaged pre-flare values of the total helicity, $\hvmean$, in the approximate range of $0.12\times10^{43} \, \si{{Mx}^{2}}$ to $59\times10^{43} \, \si{{Mx}^{2}}$ (see \gm{Col.} 3 of \href{t2}{Table~\ref{t2}}). For all the ARs, we find that $\hvmean$ is dominated by the contribution of $\hpjmean$ (\gm{Col.} 5 of \href{t2}{Table~\ref{t2}}). We find differences, however, regarding the amount to which $\hpjmean$ exceeds $\hjmean$ (\gm{Col.} 4 of \href{t2}{Table~\ref{t2}}). In particular, we find that $\hpjmean$ exceeds $\hjmean$ by a factor of $\lesssim10$ for ARs that produce large eruptive flares (with the sole exception of AR 11890), whereas it is $\gtrsim10$ for all the ARs which were the source of the large confined flares. The time-averaged pre-flare values of the total magnetic energy, $\Etotmean$ (\gm{Col.} 6 of \href{t2}{Table~\ref{t2}}), are in the approximate range $(1-18)\times10^{33}$ erg. Furthermore, the time-averaged pre-flare values of the free energy, $\Efreemean$ (last column of \href{t2}{Table~\ref{t2}}), span the approximate range $(0.9-20)\times10^{32}$ erg. \gm{The aforementioned trends of the pre-flare averages of the extensive quantities show that they are not discriminative of the flare type (confined or eruptive).}

\gm{The time-averaged pre-flare values of the normalized helicity, $\hvnormmean$, are in the approximate range (0.4 -- 9) $\times10^{-2}$, and are not distinctly different for the two types of flares. The situation is different for the time-averaged pre-flare values of the other intensive quantities,  $\hjprimemean$, $\Efprimemean$ and $\hjnormmean$, however, which seem to be discriminative in regard of whether the flare is eruptive or confined. In particular, for the majority of the eruptive flares under study, we find $\hjprimemean \gtrsim0.1$, $\Efprimemean \gtrsim0.2$, and $\hjprimemean \gtrsim0.005$, i.e., tendentially higher than for the majority of confined flares under study.}

\section{Discussion}\label{Discussion}
We investigated the magnetic helicity and energy budgets of ten flare-productive active regions around the time of occurrence of large flares (GOES class M1.8-X9.3). Our goal was to identify characteristic pre-flare magnetic energy and helicity budgets and to inquire regarding a possible relation to the type of flare produced (confined or eruptive). Thus, we computed relative magnetic helicities by applying a finite volume method. The helicity and energy computations are based on 3D coronal magnetic fields that are reconstructed based on a nonlinear force-free approach (see \href{ss:helicity}{\gm{Sect.}~\ref{ss:helicity}} for details).

Dedicated studies have been performed using solar-like MHD simulations, or observation-based force-free modeling of individual solar ARs. Those pioneering studies already pointed toward the potential of intensive quantities, and in particular that of the non-potential helicity ratio, to indicate the eruptive character of a flaring AR. In our work, we go a step further and systematically analyze in detail the time evolution of extensive ($\hv$, $\hj$, $\hpj$, $\Etot$, $\Efree$)  as well as intensive ($\hjprime$, $\Efprime$, $\hjnorm$, $\hvnorm$) quantities, during an extended time window before and around the occurrence of large solar flares.

Our results on the extensive quantities agree with past studies of the same ARs (if existing; see description of individual ARs in \href{ss:Extensive}{Sect.~\ref{ss:Extensive}}.We find no indication for extensive quantities, such as the total and free magnetic energy or the relative helicities, to be significantly different in the pre-flare corona of large flares that are eruptive or confined (see values listed in \href{t2}{Table~\ref{t2}}). This is in agreement with earlier findings in that regard, based on either MHD simulations \citep{2017A&A...601A.125P} or observation-based studies \citep{2019A&A...628A..50M,2019ApJ...887...64T}. Also in agreement with these earlier studies, we find that the normalized total helicity, $\hvnorm$, is not distinctly different for ARs hosting large eruptive or confined flares (see fifth column in \href{t3}{Table~\ref{t3}}). However, the situation is different for the other intensive quantities (\href{t3}{Table~\ref{t3}}), as discussed in the following.

We find that most of the analyzed events fall in one of two categories. One which exhibits time-averaged pre-flare values of $\hjprimemean>0.1$, $\Efprimemean>0.2$, and $\hjnormmean>0.005$. Four out of five eruptive flares fall into this category. Another category of ARs exhibits time-averaged pre-flare values of $\hjprimemean\lesssim0.1$, $\Efprimemean\lesssim0.1$, and $\hjnormmean\lesssim0.002$. Four out of five confined flares hosted by these ARs fall into this category. Only two ARs of our sample cannot be unambiguously grouped to either of the two aforementioned categories, because they show unexpected values in at least one of the intensive quantities listed above. In the following, we provide possible explanations for those two ARs.

AR 11890 exhibits time-averaged pre-flare levels of $\hjprimemean\lesssim0.1$ and $\Efprimemean\lesssim0.1$, that are typical for the confined flares in our sample. A possible explanation for the CME having occurred anyways is its relative position with respect to the AR center, as it originated from the peripheries of AR 11890 (consistent with the findings in \cite{2018EGUGA..20.5038B} about the flare locations within ARs). Another example of this kind is AR 12192, which clearly falls into the category of non-eruptive ARs. Though the majority of flares hosted by that AR were confined (all M- and X-class flares except one) a narrow CME (not studied here) was produced during one M-class flare which also originated from the periphery of the AR \citep[e.g.,][]{2015ApJ...801L..23T}. This indicates that the helicity and free energy ratio might only be indicative for eruptive vs. confined flare activity of the AR core fields, but not for flares taking place in the periphery of ARs.

The largest flare produced by AR 11166, and analyzed in our work, was confined, yet does the AR exhibit a time-averaged pre-flare value of $\hjprime$, typical for ARs that host large eruptive flares. The reason, now in retrospect, is explained easily. First, the confined nature of the X1.5 flare was actually exceptional, as the CME-productivity of the AR was in general not low  (11 out of 20 M- and C-class flares had an associated CME; see \cite{2011AGUFMSH13B1965Y}). This highlights the importance to investigate the time evolution of a solar AR during its entire disk passage, i.e., not only around individual flare events and possibly already starting at the time of AR emergence, when trying to assess its eruptive potential. 
Second, high values for the time-averaged pre-flare $\hjprime$ were found due to significant variations of the magnitude of the total helicity during the considered pre-flare time window of five  hours (see {\href{H_Hpj}{\gm{Fig.}~\ref{H_Hpj}(j)}}). In comparison, for the immediate pre-flare corona, covering about one hour prior to the start of the confined X1.5 flare, one would indeed find $\hjprime<0.1$, i.e., as for the other large confined flares studied. This highlights the need to better understand the (pre-flare) time scales relevant for flare processes.

Finally, we stress that also the most promising intensive quantities,$\hjprime$, $\Efprime$, and $\hjnorm$, are only suggestive of whether or not an AR may produce an eruptive large flare at some later stage in time. However, they are not indicative, whether or not the \textit{next} occurring flare will be confined or eruptive. As an example, AR 12673 exhibited pre-flare values prior to the analyzed eruptive X9.3 flare that are characteristic for a CME-productive AR. However, it also produced a major confined X2.2 flare a few hours earlier, for which the pre-flare averages were not drastically different, but partly even larger than prior to the eruptive flare.

\section{Summary and Conclusions}
We performed  a  first  comprehensive  study of a sample of ten ARs, for a detailed analysis of the time evolution of the coronal magnetic energy and helicity around the time of large flares (GOES class $\geq$M1.0). We computed the relative magnetic helicity using the finite volume method of \cite{2011SoPh..272..243T} based on NLFF modeling \citep{2010A&A...516A.107W}. 
Our major findings are as follows:
 \begin{enumerate}[label={\roman*)}]
  \item The extensive quantities ($\Etot$, $\Efree$, $\hv$, $\hj$, $\hpj$) cover a broad range of magnitudes with their pre-flare \gm{levels, being} not discriminative regarding the eruptive potential of the flare-producing AR.
  \item Among the intensive quantities, $\hvnorm$, is not distinctly different for ARs that \gm{produce an eruptive} or confined flare. However, the other intensive quantities, such as the helicity ratio ($\hjprime$), the energy ratio ($\Efprime$) and the normalized non-potential helicity ($\hjnorm$) are able to distinguish the pre-flare corona of ARs that possibly produce a large eruptive or confined flare. We find characteristic pre-flare values of $\hjprimemean>0.1$, $\Efprimemean>0.2$, and $\hjnormmean>0.005$ for ARs that produce large eruptive flares. For ARs that produce mostly large flares that are confined, we find $\hjprimemean\lesssim0.1$, $\Efprimemean\lesssim0.1$, and $\hjnormmean\lesssim0.002$.
  \item All helicity related intensive quantities studied here show a pronounced response during the flare if it was associated with a CME. 
\end{enumerate}

These major findings support that the combined analysis of intensive quantities (energy ratio and helicity ratio) allows it to successfully characterize the potential of an AR to produce a CME. Such an analysis, however, does not allow it to unambiguously determine the flare type of the next occurring (upcoming) flare, i.e., whether or not the next occurring flare will be eruptive or confined. For instance, pre-flare values characteristic for CME-productivity are found for both, a preceding confined and a subsequent eruptive X-class flare in AR 12673. Therefore, besides the coronal pre-requisites being met for CME-productivity, the reasons for eruptive flares to occur must also be sought for elsewhere, among others, including the particular location within an AR where a flare occurs and the strength and orientation of the overlying field \citep{Wang_2007, 2018EGUGA..20.5038B}. In exceptional cases (ARs 11890 and 12192), we have seen that a CME may erupt if it emerges from the periphery of an AR, even though the helicity and energy ratio of the underlying active region corona are untypical for CME-productive ARs. Therefore, we assume our measures to be representative mainly for the AR cores, where strong currents reside, and thus the gross of free magnetic energy and current-carrying (non-potential) helicity resides.

As a conclusion, we may say that an AR exhibiting coronal values of $\hjprimemean\lesssim0.1$, $\Efprimemean\lesssim0.1$, and $\hjnormmean\lesssim0.002$ is unlikely to produce large eruptive flares from its AR core. On the other hand, we may safely assume that if the active-region corona exhibits values of $\hjprimemean>0.1$, $\Efprimemean>0.2$, and $\hjnormmean>0.005$, the underlying AR is likely to produce a large eruptive flare subsequent in time.

\begin{acknowledgements} 
We thank the anonymous referee for valuable suggestions to improve the clarity of the manuscript. J. K. T. and M. G. were supported by the Austrian Science Fund (FWF): P31413-N27. SDO data are courtesy of the NASA/SDO AIA and HMI science teams. This article profited from discussions during the meetings of the ISSI International Team Magnetic Helicity in Astrophysical Plasmas.
\end{acknowledgements}

\bibliographystyle{aa} % style aa.bst
\bibliography{main} % your references Yourfile.bib

\clearpage
\begin{appendix}

\section{Quality metrics}\label{appendixA}
To determine the force-freeness of a NLFF modeled field, the current-weighted average of the sine of the angle between the modeled magnetic field and the electric current density \citep[see,][]{2000ApJ...540.1150W,2006SoPh..235..161S} is calculated as,
\begin{eqnarray}
 \sigma_{J}=\sin{\theta_{J}}=\left(\sum_{i}^{}\frac{|\vec{J}_{i}\times\vec{B}_{i}|}{B_{i}}\right)/ \sum_{i}^{} J_{i},
\end{eqnarray}
where $\thetaj=\sin^{-1}{\sigma_{J}}$ is then, the current-weighted angle between the $\Bvec$ and $\Jvec$. For a completely force-free field, $\sigma_{J}=0$, since $\thetaj=0$.

In order to quantify the  solenoidality (divergence freeness) of a magnetic field, \cite{2013A&A...553A..38V} introduced a parameter, $\Edivprime$, which expresses the non-solenoidal fraction of the total energy based on the decomposition
\begin{eqnarray}
\begin{split}
	\Etot & = \frac{1}{8\pi}\int_V\textbf{B}^2\;{\rm ~d}V\\
	& =\Eps+\Epns+\Ejs+\Ejns+\Emix. \label{eq:edive}
\end{split}
\end{eqnarray}
Here $\Eps$ and $\Ejs$ are the energies of the solenoidal components of the potential ($\Epot$) and current-carrying ($\Ej$) magnetic field, respectively, while $\Epns$ and $\Ejns$ are the energies of the corresponding non-solenoidal components. All contributing energies, except for $\Emix$, are positive definite. $\Emix$ is the energy corresponding to all cross terms (see Eq. (8) in \cite{2013A&A...553A..38V}). The energy associated with all non-solenoidal components can then be defined as
\begin{eqnarray}
	\Ediv = \Epns+\Ejns+|\Emix|. \label{eq:ediv}
\end{eqnarray}
Usually, {$\Eps>\Ejs>\Emix>\Ejns>\Epns$}
\citep[see, \eg,][]{2015ApJ...811..107D}. 
In case of a perfectly solenoidal magnetic field, one finds $\Epns=\Ejns=\Emix=0$, and therefore $\Ediv=0$.

For each AR under study, the values of $\thetaj$ and $\Edivprime$ are shown in \href{CW_theta}{\gm{Fig.}~\ref{CW_theta}} \& \href{Ediv_E}{~\ref{Ediv_E}}, respectively. NLFF model time series with enforced lower residual divergence exhibit lower values of $\Edivprime$ (see square symbols in \href{Ediv_E}{\gm{Fig.}~\ref{Ediv_E}}), at the slight expense of force-freeness (compare \href{CW_theta}{\gm{Fig.}~\ref{CW_theta}}). Average values of $\thetaj$ are below 15 degrees at most of the time instances. Also, average values of $\Edivprime$ are $\lesssim0.1$ for most time instances considered for the individual ARs. We remind the reader here that only NLFF solutions exhibiting values of $\Edivprime<0.1$ are qualifying for subsequent helicity computation. All others are excluded from further analysis.

\begin{figure*}[htp]
    \captionsetup{width=.9\linewidth}
    \appfig{CW_theta}{.pdf}
    \caption{Quality metric $\thetaj$ at each time instant for the 10 ARs under study. The left column shows ARs that produce large eruptive flares, the right column shows ARs that produce large confined flares. Squares (bullets) show $\thetaj$ of the solutions based on enhanced (standard) divergence freeness. The gray shaded areas mark the spread of $\thetaj$, and the black curve indicate the average values. The vertical bar marks the impulsive flare phase.}
    \label{CW_theta}
\end{figure*}

\begin{figure*}[htp]
    \captionsetup{width=.9\linewidth}
    \appfig{Ediv_E}{.pdf}
    \caption{Quality metric $\Edivprime$ at each time instant for the 10 ARs under study. The left column shows ARs that produce large eruptive flares, the right column shows ARs that produce large confined flares. Squares (bullets) show $\Edivprime$ of the solutions based on enhanced (standard) divergence freeness. The gray shaded areas mark the spread of $\Edivprime$, and the black curve indicate the average values. The vertical bar marks the impulsive flare phase.}
    \label{Ediv_E}
\end{figure*}

\end{appendix}

\end{document}